# Kappa-deformed random-matrix theory based on Kaniadakis statistics

A.Y. Abul-Magd[1,2] and M. Abdel-Megeed[2]
[1]Faculty of Engineering, Sinai University, El-Arish, Egypt
[2]Faculty of Science, Zagazig University, Zagazig, Egypt

October 29, 2018

**Abstract**

We present a possible extension of the random-matrix theory, which is widely used to describe spectral fluctuations of chaotic systems. By considering the Kaniadakis non-Gaussian statistics, characterized by the index $\kappa$ (Boltzmann-Gibbs entropy is recovered in the limit $\kappa \to 0$), we propose the non-Gaussian deformations ($\kappa \neq 0$) of the conventional orthogonal and unitary ensembles of random matrices. The joint eigenvalue distributions for the $\kappa$-deformed ensembles are derived by applying the principle maximum entropy to Kaniadakis entropy. The resulting distribution functions are base independent as they depend on the matrix elements in a trace form. Using these expressions, we introduce a new generalized form of the Wigner surmise valid for nearly-chaotic mixed systems, where a basis-independent description is still expected to hold. We motivate the necessity of such generalization by the need to describe the transition of the spacing distribution from chaos to order, at least in the initial stage. We show several examples about the use of the generalized Wigner surmise to the analysis of the results of a number of previous experiments and numerical experiments. Our results suggest the entropic index $\kappa$ as a measure for deviation from the state of chaos. We also introduce a $\kappa$-deformed Porter-Thomas distribution of transition intensities, which fits the experimental data for mixed systems better than the commonly-used gamma-distribution.

# 1 Introduction

Quantum systems whose classical counterparts have chaotic dynamics are often modelled in terms of random-matrix theory (RMT) [1, 2]. This is the statistical theory of random matrices $H$ whose entries fluctuate as independent Gaussian random numbers. Dyson [3] showed that there are three generic ensembles of random matrices, defined in terms of the symmetry properties of the Hamiltonian. The most popular of these is the Gaussian orthogonal ensemble (GOE) which successfully describes many time-reversal-invariant quantum systems whose classical counterparts have chaotic dynamics. Systems without time-reversal symmetry are modelled by the Gaussian unitary ensemble (GUE). Balian [4] derived the weight functions for the random-matrix ensembles from Jaynes' principle of maximum entropy (MaxEnt) [5]. He applied the conventional Shannon entropy to ensembles of random matrices with distribution $P(H)$ and maximized it under the constraints of normalization of $P(H)$ and fixed mean value of $\mathrm{Tr}\big(H^{\dagger}H\big)$, where $H^{\dagger}$ is the Hermitian conjugate of $H$. The latter constraint ensures basis independence, which is a property of the trace of a matrix. The resulting expression

$$P(H) \propto \exp\left[-\eta \mathrm{Tr}\left(H^{\dagger}H\right)\right] \tag{1}$$

is supposed to account for all possible level statistics of various chaotic systems.

For most systems, however, the phase space is partitioned into regular and chaotic domains. These systems are known as mixed systems. Attempts to generalize RMT to describe such mixed systems are numerous; for a review please see [2]. MaxEnt can contribute to the generalization of RMT by either introducing additional constraints or modifying the entropy. An example of the first approach is the work of Hussein and Pato [7], who use MaxEnt to construct "deformed" random-matrix ensembles by imposing different constraints for the diagonal and off-diagonal elements. The second approach to include mixed systems in RMT is to apply MaxEnt to entropic measures other than Shannon's. Indeed, there is no systematic way of deriving the "right" entropy for a given dynamical system. Different entropic forms have been introduced instead of Shannon's entropy, which would generalize the classical statistical mechanics of Boltzmann and Gibbs. During the last two decades a great interest has arisen in the non-extensive Tsallis statistical mechanics [8], and, more recently, on the Kaniadakis extensive generalized power-law statistics [9] motivated by various restrictions to the applicability of classical, extensive statistical mechanics. This has been accomplished with the introduction of an appropriate deformed logarithm [**?**]. Recently, the Tsallis formalism has been applied to include systems with mixed regular-chaotic dynamics in a nonextensive generalization of RMT [10, 11, 12, 13, 14, 15]. The matrix-element distribution $P(H)$ is obtained by extremizing Tsallis' $q$-entropy rather than Shannon's, but again subject to the same constraints of normalization and constant expectation value of $\mathrm{Tr}\big(H^{\dagger}H\big)$. This distribution function, as well as the distributions to be obtained for other generalized entropies, depends on the matrix elements in the combination $\mathrm{Tr}\big(H^{\dagger}H\big)$. Therefore, the random-matrix ensembles described by these distribution functions are base independent.

Another nonextensive approach to RMT [16] replaces Shannon's entropy in the conventional theory by Kaniadakis' $\kappa$-entropy [9, 17, 18, 19]. The Kaniadakis entropy is maximized again under the constraints of normalization of $P(H)$ and fixed mean value of $\mathrm{Tr}\big(H^{\dagger}H\big)$. Clearly, the resulting distribution function depends on the matrix elements in the combination $\mathrm{Tr}\big(H^{\dagger}H\big)$ as in the case with Tsallis' entropy. Therefore, the emerging random-matrix ensemble is also base independent.

In this paper, by following our previous results [16], we adopt the Kaniadakis non-Gaussian statistics and study the effects of the $\kappa$-generalization for RMT, especially what concerns its effects on NNS of the eigenvalues of GOE and GUE of random matrices. We also investigate the influence of these non-gaussian effects on transitions in the mixed system out of the chaotic phase.

This paper is organized as follows. A brief review of Kaniadakis statistics is presented in Sec. 2. In Sec. 3, we consider the connection between the $\kappa$-generalization of RMT and the transition of a quantum system out of the chaotic phase. Section 4 presents a $\kappa$-generalization of the Wigner surmise which is a good approximation for the spacing distribution of chaotic systems. A comparison between the predictions of the $\kappa$-deformed Wigner surmise with the results of a numerical experiment is given in Sect. 5. Section 6 presents a formula for the transition intensities which is shown to be more consistant with experimental data than the gamma distribution which is commonly used in data analysis. We summarize our main conclusions in Sect. 7.

## 2 Kaniadakis' entropy

In this section, we present the basic elements of Kaniadakis' $\kappa$-deformed framework introduced in [9]. The Kaniadakis non-gaussian statistics in turn is characterized by the $\kappa$-entropy that emerges naturally in the framework of the so-called kinetic interaction principle, which underlies the non-linear kinetics in particle systems. This principle fixes the expression of the Fokker-Planck equation describing the kinetic evolution of the system and imposes the form the generalized entropy associated with the system. The structure of the ensuing $\kappa$-deformed statistical mechanics has striking similarity with that of special relativity, suggesting that it might be relevant for the self-consistent formulation of relativistic statistical theory [18, 19] and relativistic Boltzmann kinematics [20] with application to, e.g., relativistic distribution of fluxes of cosmic rays particle systems [19], the formation of a quark-gluon plasma [21], the relativistic nuclear equation of state [22], the relativistic gas in an electromagnetic field [23], and the relaxation of relativistic plasmas under the effect of wave-particle interactions [24]. The $\kappa$-deformed statistics have also been successfully applied in several branches of physics. Examples are the nonlinear kinetics [20], and the H-theorem from a generalization of the chaos molecular hypothesis [25], the sensitivity to initial conditions and the entropy production in the logistic map [26], personal income distribution for Germany, Italy and UK [27], the distribution of stellar rotational velocity for stars belonging to various classes [28] and systems of interacting atoms and photons [29].

From the mathematical point of view, Kaniadakis' statistics is based on the $\kappa$-deformed exponential [9], which is defined by

$$\exp_\kappa(x) = \left(\sqrt{1+\kappa^2 x^2} + x\right)^{1/\kappa} = \exp\left(\frac{1}{\kappa}\text{arcsinh}\;\kappa x\right). \qquad (2)$$

The properties of the function $\exp_\kappa(x)$ have been considered extensively in the literature. We recall briefly that in the $\kappa \to 0$ limit the function $\exp_\kappa(x)$ reduces to the ordinary exponential, i.e. has the properties $\exp_0(x) = \exp(x)$. For $x \to 0$, it behaves very similarly with the ordinary exponential independently on the value of $\kappa$. It is remarkable that the first three terms of the Taylor expansion

$$\exp_\kappa(x) = 1 + x + \frac{x^2}{2} + \left(1 - \kappa^2\right)\frac{x^3}{3!} + ... \tag{3}$$

are the same as in the ordinary exponential. On the other hand, the most interesting property of $\exp_\kappa(x)$ for the applications in statistics is the power-law asymptotic behavior

$$\exp_\kappa(x) \underset{x\to\pm\infty}{\sim} |2\kappa x|^{\pm 1/|\kappa|}. \tag{4}$$

The corresponding logarithmic function $\ln_\kappa(x)$ is defined as the inverse function of $\exp_\kappa(x)$, namely $\ln_\kappa(\exp_\kappa(x)) = \exp_\kappa(\ln_\kappa(x)) = x$, and is given by

$$\ln_\kappa(x) = \frac{x^\kappa - x^{-\kappa}}{2\kappa}. \tag{5}$$

Kaniadakis' entropy associated with the $\kappa$-statistics is obtained by replacing the logarithm in the expression for the standard Shannon's entropy by the $\kappa$-logarithm. For a distribution function $f$, the $\kappa$-entropy is given by

$$S_{\mathrm{K}} = -\int d\Gamma f \ln_\kappa f \tag{6}$$

where $d\Gamma$ is the phase-space volume element. Maximizing the $\kappa$-entropy with condition of fixed mean value of a Hamiltonian $h$ yields

$$P(H) = \frac{1}{z_\kappa}\exp_\kappa(-\beta h), \tag{7}$$

where $\beta$ is a Lagrange multiplier, and $z_\kappa$ is a moralization factor. We see that the Gaussian probability curve is replaced by the characteristic power-law behavior of Kaniadakis framework and, as expected, the limit $\kappa = 0$ recovers the exponential result.

# 3 The $\kappa$-deformed RMT

In this section, we consider a possible generalization of RMT based on an extremization of Kaniadakis' $\kappa$-entropy. We obtain the Kaniadakis entropy for the matrix-element probability distribution function using Eqs. (5) and (6). Explicitly, it reads

$$S_{\mathrm{K}}\left[\kappa, P_{\mathrm{K}}(\kappa, H)\right] = -\frac{1}{2\kappa}\int dH \left(\frac{\alpha^\kappa}{1+\kappa}\left[P_{\mathrm{K}}(\kappa, H)\right]^{1+\kappa} - \frac{\alpha^{-\kappa}}{1-\kappa}\left[P_{\mathrm{K}}(\kappa, H)\right]^{1-\kappa}\right) \tag{8}$$

where $\kappa$ is a parameter with value between 0 and 1. The case of $\kappa = 0$, corresponds to the Shannon entropy. Here, $\alpha$ is a real positive parameter. Kaniadakis has considered two choices of $\alpha$, namely $\alpha = 1$ and $\alpha = Z$, where $Z$ is a generalized partition function. We here adopt the second choice. The matrix-element distribution $P_{\rm K}(\kappa, H)$ is obtained by extremizing the $\kappa$-entropy subject to the condition of fixed mean value of $\mathrm{Tr}\big(H^\dagger H\big)$. One has to obtain the extremum of the functional

$$F_{\rm K} = S_{\rm K} - \eta_{\rm K} \int dH \; P_{\rm K}(\kappa, H) \mathrm{Tr}\left(H^T H\right), \tag{9}$$

where $\eta_{\rm K}$ is a Lagrange multiplier. One arrives to the following distribution

$$P_{\rm K}(\kappa, H) = \frac{1}{Z} \exp_\kappa \left[-\eta_{\rm K} \mathrm{Tr}\left(H^T H\right)\right], \tag{10}$$

where

$$Z = \int dH \;\; \exp_\kappa \left[-\eta_{\rm K} \mathrm{Tr}\left(H^T H\right)\right]. \tag{11}$$

Accordingly, the $\kappa$-deformed matrix-element distribution $P_{\rm K}(\kappa, H)$ is obtained by replacing the exponential function in Eq. (1) by the $\kappa$-exponential given by identity (2). One arrives to the following distribution

$$P_{\rm K}(\kappa, H) = \frac{1}{Z_{\;\kappa}} \exp_\kappa \left[-\eta_{\rm K} \mathrm{Tr}\left(H^\dagger H\right)\right], \tag{12}$$

in which

$$Z_\kappa = \int dH \;\; \exp_\kappa \left[-\eta_{\rm K} \mathrm{Tr}\left(H^\dagger H\right)\right] bv. \tag{13}$$

is a normalization factor defined by

$$Z_\kappa = 2^{-\beta N(N-1)/4} \int d^n\mathbf{r} \exp\left(-\frac{1}{\kappa} \mathrm{arcsinh}\; \kappa \eta_{\rm K} r^2\right) = 2^{-\beta N(N-1)/4} \left(\frac{\pi}{\eta_{\rm K}}\right)^{n/2} \frac{\Gamma_\kappa\left(\frac{n}{2}\right)}{\Gamma\left(\frac{n}{2}\right)} \tag{14}$$

for $\kappa < 2/n$. Here we define a $\kappa$-deformed gamma function by

$$\Gamma_\kappa\left(n\right) = \int_0^\infty x^{n-1} \exp\left(-\frac{1}{\kappa} \mathrm{arcsinh}\; \kappa x\right) dx = \frac{1}{\left(2\kappa\right)^{1+n}} \frac{\Gamma\left(n\right) \Gamma\left(\frac{1}{2\kappa} - \frac{n}{2}\right)}{\Gamma\left(1 + \frac{1}{2\kappa} + \frac{n}{2}\right)}. \tag{15}$$

A similar result is obtained by Kaniadakis in evaluating an analogous integral that appears in his treatment of Brownian particles (Eq. (71) of Ref. [9]).

We now calculate the joint probability density for the eigenvalues of the Hamiltonian $H$. With $H = U^{-1} E U$, where $U$ is the global unitary group, we introduce the elements of the diagonal matrix of eigenvalues $E = \mathrm{diag}(E_1, \cdots, E_N)$ of the eigenvalues and the independent elements of $U$ as new variables. Then the volume element $dH$ has the form

$$dH = \left|\Delta_N\left(E\right)\right| dE d\mu(U), \tag{16}$$

where $\Delta_N(E) = \prod_{n>m}(E_n - E_m)$ is the Vandermonde determinant and $d\mu(U)$ the invariant Haar measure of the unitary group [1, 2]. The probability density $P_{\mathrm{K}}(\kappa, H)$ depends on $H$ through $\mathrm{Tr}\left(H^{\dagger}H\right)$ and is therefore invariant under arbitrary rotations in the matrix space. Integrating over the "angular variables" $U$ yields the joint probability density of eigenvalues in the form

$$P_{\mathrm{K}}(\kappa; E_1, ..., E_N) = C \prod_{n>m} |E_n - E_m|^{\beta} \exp_{\kappa}\left[-\eta_{\mathrm{K}} \sum_{i=1}^{N} E_i^2\right]. \tag{17}$$

where $C$ is a normalization constant.

# 4 The $\kappa$-deformed Wigner surmises

The calculation of NNS distributions starting from the matrix-element probability density function (1) is a formidable task. In fact, RMT does not provide simple analytical expressions for NNS distributions. There are several elaborate approaches to evaluate these distributions [1]. The author is not aware of application of these results to the analysis of experimentally observed or numerically calculated discrete data.

In many cases, the empirical data are actually compared to the so-called Wigner surmises, which correspond to the NNS distributions of ensembles of $2 \times 2$ matrices. For the three Gaussian ensembles, the Wigner surmises are given by

$$P_{\beta}(s) = a_{\beta} s^{\beta} \exp\left(-b_{\beta} s^2\right), \tag{18}$$

where $\beta = 1, 2,$ and 4 for GOE, GUE and GSE, respectively. The factors $a_{\beta}$ and $b_{\beta}$ are obtained from the normalization condition and the requirement of a unit mean spacing,

$$\int_0^{\infty} P_{\beta}(s) ds = 1, \;\; \int_0^{\infty} s P_{\beta}(s) ds = 1. \tag{19}$$

For GOE, $a_1 = \pi/2$ and $b_1 = \pi/4$ while, for GUE, $a_2 = 32/\pi$ and $b_2 = 4/\pi$. The Wigner surmises present accurate approximation to the exact results for the case of large $N$.

We consider the $\kappa$-deformed generalization of the Wigner surmises hoping that they present an accurate approximation for $\kappa$-deformed ensembles of large $N$, which has an equal success to that of the standard Wigner surmises. For this purpose we follow the arguments given in the work of Carvalho et al. [28]. We replace the exponential function in Eq. (18) by the $\kappa$-deformed exponential in Eq. (2)to obtain

$$P_{\mathrm{K}}(\beta, \kappa; s) = a(\beta, \kappa) s^{\beta} \exp_{\kappa}\left[-b(\beta, \kappa) s^2\right]. \tag{20}$$

Again the factors $a(\beta, \kappa)$ and $b(\beta, \kappa)$ are obtained from the normalization and unit-mean-spacing conditions (19) so that

$$a(\beta, \kappa) = \frac{2 b^{\frac{\beta+1}{2}}(\beta, \kappa)}{\Gamma_{\kappa}\left(\frac{\beta+1}{2}\right)} \tag{21}$$

and

$$b(\beta,\kappa) = \left[\frac{\Gamma_\kappa\left(\frac{\beta}{2}+1\right)}{\Gamma_\kappa\left(\frac{\beta+1}{2}\right)}\right]^2. \tag{22}$$

We can easily sea that $a(1,0) = \pi/2$ and $b(1,0) = \pi/4$ while $a(2,0) = 32/\pi$ and $b(0,\kappa) = 4/\pi$. Therefore, as $\exp_0(x) = \exp(x)$, the proposed $\kappa$-deformed Wigner surmises $P_{\mathrm{K}}(\beta,\kappa;s)$ tend to the conventional ones as $\kappa \to 0$. Using the small argument expansion for $\exp_\kappa(x)$ (3), Eq. (20) yields

$$P_{\mathrm{K}}(\beta,\kappa;s) = a(\beta,\kappa)s^\beta\left[1 + O\left(s^2\right)\right], \tag{23}$$

suggesting that the $\kappa$-deformed Wigner surmises have the same power-law level repulsion as the original ones. On the other hand, the large-argument behavior of $\exp_\kappa(x)$ in Eq. (4) implies the following power-law asymptotic behavior of $P_{\mathrm{K}}(\beta,\kappa;s)$

$$P_{\mathrm{K}}(\beta,\kappa;s) \propto s^{1-2/\kappa}, \text{ for } s \gg \frac{1}{\sqrt{\kappa b(\beta,\kappa)}} \approx \sqrt{\frac{2}{\pi\kappa}}. \tag{24}$$

The mean square spacing of the distribution (20), $\langle s^2\rangle = \int_0^\infty s^2 P_\beta(s)ds$, is given by

$$\langle s^2\rangle = \frac{(1+\beta)\Gamma\left(\frac{1}{2\kappa}-\frac{\beta}{4}-\frac{3}{4}\right)\Gamma\left(\frac{1}{2\kappa}+\frac{\beta}{4}+\frac{5}{4}\right)}{4\kappa\Gamma\left(\frac{1}{2\kappa}-\frac{\beta}{4}-\frac{1}{4}\right)\Gamma\left(\frac{1}{2\kappa}+\frac{\beta}{4}+\frac{7}{4}\right)}. \tag{25}$$

It is reasonable to suggest that the condition of fixed mean value of $\mathrm{Tr}\left(H^\dagger H\right)$ implies that $\langle s^2\rangle$ is finite. This restricts the range of allowed values of $\kappa$ to the following

$$0 \le \kappa < \frac{2}{3+\beta}. \tag{26}$$

# 5 NNS distribution for mixed systems

Bohigas et al. [32] put forward a conjecture (strongly supported by accumulated numerical evidence) that the energy levels of a typical time-reversal invariant chaotic systems have the same fluctuation statistics as that of GOE. In particular, the NNS distribution is given by the Wigner surmise (18) with $\beta = 1$

$$P_{\mathrm{GOE}}(s) = \frac{\pi}{2}s\exp\left(-\frac{\pi}{4}s^2\right) \tag{27}$$

In the nearly ordered regime, mixing of quantum states belonging to adjacent levels can be ignored and the energy levels are uncorrelated. The level-spacing distribution function obeys the Poissonian

$$P_{\mathrm{Poisson}}(s) = \exp(-s) \tag{28}$$

Systems with mixed regular-chaotic dynamics have NNS distribution intermediate between the Wigner and Poisson distributions. Several formulae were suggested to interpolate between these two distributions [2].

The NNS statistics of mixed quantum systems execute a Poisson-to-Wigner transition as the underlying classical dynamics monotonically change from being completely integrable to completely chaotic. We explore in this section the relevance of the $\kappa$-deformed Wigner surmise (20) for mixed systems. We do this phenomenologically by comparing its prediction with the experimental or numerical-experimental NNS distributions for a number of mixed systems. We confine our consideration to the systems that conserve time reversal systems. Our purpose is to find out how far can we describe the Poisson-GOE transition in terms of Eq. (20) with $\beta = 1$. Explicitly, the $\kappa$-deformed Wigner surmise for Poisson-GOE transition reads

$$P_{\text{K-GOE}}(\kappa, s) = A(\kappa) s \exp_\kappa \left[ -B(\kappa) s^2 \right], \tag{29}$$

where

$$A(\kappa) = 2(1-\kappa^2)B(\kappa), \;\; B(\kappa) = \frac{\pi}{2}\left[\frac{(1-\kappa^2)\Gamma\left(\frac{1}{2\kappa}-\frac{3}{4}\right)}{8\kappa^{5/2}\Gamma\left(\frac{1}{2\kappa}+\frac{7}{4}\right)}\right]^2. \tag{30}$$

As mentioned above, the requirement of finite second moment of the spacing distribution restricts the range of variation of $\kappa$ to be

$$0 \le \kappa < \frac{1}{2}, \tag{31}$$

with $\kappa = 0$ yielding $P_{\text{GOE}}(s)$. The evolution of the distribution $P_{\text{K-GOE}}(1/2, s)$ in this interval is shown in Fig. 1. As $\kappa$ increases, the peak of $P_{\text{K-GOE}}(\kappa, s)$ moves towards small value of $s$. At $\kappa = 1/2$, the distribution has its peak at the same position as the semi-Poisson distribution

$$P_{\text{SP}}(s) = 4s\exp(-2s). \tag{32}$$

The semi-Poisson distribution was suggested to describe a narrow intermediate region between insulating and conducting regimes exemplified by the Anderson localization model [33], with the two limiting cases being described by Poisson and Wigner statistics, respectively. It was introduced to mimic new seemingly universal properties in certain classes of systems, in particular, being characteristics of the "critical quantum chaos". Figure 1 compares the limiting superstatistical distribution, $P_{\text{K-GOE}}(1/2, s)$, with the semi-Poisson distribution. We see from the figure that the two distributions are quite similar in shape. Therefore it is worth in the present context to use the semi-Poisson statistics as a reference distribution marking the limit of validity of the base-invariant random-matrix description of mixed systems. Analogous conclusion was reached in the analysis of the transition out of chaos using the superstatistical generalization of RMT [34].

In the following, we test how accurately the distribution $P_{\text{K-GOE}}(\kappa, s)$ reproduces the histogram data obtained in a number of numerical experiments.

## 5.1 Random points on fractals

Sakhr and Nieminen [35] have recently noted that the NNS distribution of random points uniformly distributed on a line is given by the Poisson distribution (28). On the other hand he NNS distribution of random points uniformly distributed on a plane is given by the Wigner surmise for GOE (27). Exploiting this observation, they introduce a model for a crossover transition between Poisson and Wigner statistics consisting of random points on a continuous family of self-similar curves with fractal dimensions between 1 and 2. This model is special since the intermediate statistics are described exactly by the Brody distribution [36]

$$P_{\mathrm{Brody}}(\omega, s) = a_\omega s^\omega \exp\left(-b_\omega s^{\omega+1}\right) \tag{33}$$

with $\omega = d_s - 1$, which is frequently used in the phenomenological analysis of mixed systems. As a concrete example, they study point processes on the family of Koch fractals in 2-dimensional space. These fractals can be thought of as the attractors of a one-parameter family of iterated function systems. They show that NNS distributions of the random points undergo a continuous transition from Poisson to Wigner statistics as the similarity dimension of the fractals $d_s$ varies from 1 (line) to 2 (plane). Their results are shown in Fig. 2 together with the best-fit curves calculated using $\kappa$-deformed Wigner surmise of Eq. (29). Figure 3 shows the absolute deviation between the theoretical and numerical-experimental curves, which is defined by

$$\Delta(\kappa) = \frac{1}{N}\sum_{i=1}^{N} \left|P_{\mathrm{K\text{-}GOE}}(\kappa, s_i) - P_{\mathrm{experiment}}(s_i)\right| \tag{34}$$

The two figures show that the agreement is satisfactory for the nearly chaotic regime (the lower panels in Fig. 2), and gradually decrease as the distributions become less chaotic to become unsatisfactory as the system approaches integrability (the upper panels in Fig. 2). If we accept an absolute deviation below $\Delta = 0.05$ as an indicator for satisfaction of the agreement with the data, which in our case corresponds to a fractional dimension (and thus Brody's parameter) nearly equal to $d_s = 1.46$ ($\omega = 0.46$). We note that the Brody distribution that best-fits the semi-Poisson distribution has a parameter $\omega = 0.41$.

The transition between integrability and chaos in this numerical experiment is monotonic. The degree of "chaoticity" is measured by the similarity dimension of the fractals $d_s$ and modelled in this paper by the Kaniadakis index $\kappa$. In many systems, however, the degree of chaoticity changes in a complicated way as a system parameter is varied monotonically, and so the energy-level statistics will not undergo a direct transition from Poisson to Wigner. In the following subsection we consider a system which undergoes a Poisson-Wigner-Poisson transition.

## 5.2 Elliptical stadium billiard

Lopac et al. [37] investigated the dynamical properties of the elliptical stadium billiard, which is a generalization of the Bunimovich circular stadium billiard [38]. In its dependence on two shape parameters $\delta$ and $\gamma$ , this system reveals a rich interplay of integrable, mixed and fully chaotic behavior. Their results for the classical dynamics confirm the existence of a large fully chaotic region surrounding the straight line $\delta = 1 - \gamma$ that corresponds to the Bunimovich billiard. The histograms of NNS distributions reported in [37] are for the elliptic billiards with $\delta = \gamma$. For these billiards, the chaotic region of the shape parameters is around $\delta = 0.5$. Fig. 4 shows the result of comparison between these histograms with the predictions of Eq. (29). The best-fit values of $\kappa$ together with the absolute deviations $\Delta$ are given in the figure. If we again accept the value of $\Delta = 0.05$ to quantify the start of unsatisfactory agreement with the data, then we consider the $\kappa$-deformed approach to be unsatisfactory outside the region of $0.4 \leq d_s \leq 0.75$. This shows the the $\kappa$-deformed RMT provides a better description for mixed systems as the evolve approaching equilibrium or departing out of it.

### 5.3 Boson systems with one- plus two-body random interaction

Random matrix ensembles (and in particular GOE) for $m$-particle system, involve interaction up to $m$-body in character and are dominated by the $m$-body interactions. In real many-particle systems such as atoms, nuclei, quantum dots etc. however, the inter-particle interaction is known to be only 1-body and 2-body in character. Hence the concept of embedded ensemble of $k$-body interaction (EGOE($k$)), was introduced [39]. Embedded ensembles are usually studied using only 2-body interactions, i.e., by allowing the single-particle energies to be degenerate. Patel et al. [40] have demonstrated that the EGOE(2) for dense interacting boson systems exhibit spectral fluctuations of GOE type. On the other hand, ensembles with only one-body interaction would show the Poisson character. Chavda et al. [41] presented numerical results for the study of the statistical properties of embedded ensembles of interacting Bosons when the interaction is taken as $H = H_1 + \lambda H_2$, so that the strength of the 2-body interaction can be varied by changing the value of $\lambda$; $H_n$ here represents the $n$-body interaction. They examined the nature of the transition in spacing distribution as the strength of the 2-body interaction is varied while keeping the nondegenerate but fixed single-particle energies They studied the nature of the spacing distribution as a function of strength $\lambda$ of the 2-body interaction. Their results are compared with with the predictions of Eq. (29) in Fig.5. We see again that In conclusion, we can see that the agreement between the histograms of the NNS distributions for bosonic ensembles and the $\kappa$-deformed Wigner surmise steadily improves as the distribution moves from Poisson type to the Wigner type.

### 5.4 Statistics of $2^+$ levels in even–even Nuclei

Originally, RMT was proposed for the purpose of describing the spectral properties of nuclear levels. Haq et al [42] have shown that the spectral fluctuation properties of nuclear resonances near nucleon emission threshold are in very good agreement with the predictions of GOE. On the other hand, NNS distributions of nuclear levels near the ground state lie between the Poisson and Wigner distributions [43]. Abul-Magd et al. [44] have collected the empirical information on low–lying levels of even–even nuclei with spin-parity $2^+$, published in Nuclear Data Sheets until July 2002. To obtain statistically relevant samples, the nuclei are grouped into classes defined by the ratio $R_{4/2}$ of the excitation energies of the first $4^+$ and $2^+$ levels. This ratio serves as a measure of collectivity in nuclei. Analyses done in Ref. [44] show that the NNS distribution for each group vary strongly with $R_{4/2}$ and takes a more regular shape in nuclei that have one of the dynamical symmetries of the interacting Boson model [45]. Figure 6 shows the result of comparison between the histograms for each group with the predictions of Eq. (29). The best-fit values of $\kappa$ together with the absolute deviations $\Delta$ are given in the corresponding windows. As seen in Fig. 6, the entropic parameter $\kappa$ is indeed dependent on $R_{4/2}$. It takes large value at $R_{4/2} = 2.0$, 2.5, and 3.3, which respectively correspond to the U(5), SO(6), and SU(3) dynamical symmetries of the interacting Boson model.

# 6 Transition intensity distribution for mixed systems

Matrix elements of transition operator probe the system's wave functions so that their statistical fluctuations provide additional information. In chaotic systems, the reduced transition probabilities follow the Porter-Thomas distribution [6]. The probability $B_{if}$ of a transition from the initial configuration $|i\rangle$ to the final configuration $|f\rangle$ is given by

$$B_{if} = |W_{if}|^2 , \tag{35}$$

where$W_{if}$ are the matrix elements of the transition operator in a special basis. If the transition operator conserves time reversibility, the matrix elements $W_{if}$ are real. For a chaotic system with time reversal symmetry, it is reasonable to assume that $W_{if}$ are identically-distributed Gaussian random variable. This entails that the transition intensities can be represented by a random variable that takes the values

$$y_{if} = \frac{B_{if}}{\langle B_{if} \rangle} \tag{36}$$

where $\langle B_{ij} \rangle$ is a suitably defined local average value [46], and has a Porter-Thomas distribution

$$P_{\mathrm{PT}}(y) = \sqrt{\frac{\eta}{\pi y}} e^{-\eta y}. \tag{37}$$

This is a $\chi^2$-distribution of one degree of freedom. The parameter $\eta$ is defined by the requirement that $\langle y \rangle = 1$, and is equal to 1/2. A more elaborate derivation of the Porter-Thomas distribution is given by Barbosa et al. [47].

As the system becomes more regular, the transition probabilities deviate from the Porter-Thomas distribution. To account for these deviations, Alhassid and Novoselsky [48] suggested that the transition widths in mixed system may be analyzed in terms of a $\chi^2$-distribution with $\nu$ degrees of freedom

$$P_{\chi^2}(y,\nu) = \frac{1}{2^{\nu/2}\Gamma\left(\frac{\nu}{2}\right)} y^{\nu/2-1} e^{-\nu y}. \tag{38}$$

When $\nu = 1$, This distribution coincides with the Porter-Thomas distribution. The $\kappa$-deformed distribution does not fit well the empirical distributions but consists better with the observed number of weak transitions as compared with the Porter-Thomas distribution (see, e.g. [46, 47, 49]). When the empirical distributions are expressed as functions of the $\log_a y$ have peaks at $\log_a y < 0$ while all the $\chi^2$ distributions are peaked at $\log_a y = 0$ for all values of $a$ and $\nu$. We show in the present paper that the $\kappa$-deformed RMT provides us with a generalization of the Porter-Thomas distribution, which is more suitable for the analysis empirical data.

## 6.1 The $\kappa$-deformed Porter-Thomas distribution

We now derive the $\kappa$-deformed generalization of the Porter-Thomas distribution. We do this by replacing the exponential function in (37) by a $\kappa$-deformed exponential. One then obtains

$$P_{\text{K-PT}}(y) = \frac{c(\kappa)}{\sqrt{\kappa}} \exp_\kappa\left[-d(\kappa)y\right] \tag{39}$$

where the quantities $c(\kappa)$ and $d(\kappa)$ are obtained from the normalization condition and the requirement that $\langle y \rangle = 1$, which yield

$$c(\kappa) = \frac{2\kappa\Gamma\left(\frac{1}{2\kappa}+\frac{5}{4}\right)}{\Gamma\left(\frac{1}{2\kappa}-\frac{1}{4}\right)} d(\kappa), \tag{40}$$

and

$$d(\kappa) = \frac{\Gamma\left(\frac{1}{2\kappa}-\frac{3}{4}\right)\Gamma\left(\frac{1}{2\kappa}+\frac{5}{4}\right)}{4\kappa\Gamma\left(\frac{1}{2\kappa}-\frac{1}{4}\right)\Gamma\left(\frac{1}{2\kappa}+\frac{7}{4}\right)}. \tag{41}$$

We compare in Fig. 7 the evolution of $P_{\text{K-PT}}(\ln y)$ during the transition from chaotic to regular dynamics by varying the tuning parameter $\kappa$ from 0 (GOE) to 0.6 (almost regular) with a corresponding evolution of $P_{\chi^2}(\ln y, \nu)$ where $\nu$ varies between 1 and 0.1. Of special interest is the fact that the maximum of $P_{\chi^2}(\ln y, \nu)$ occurs at $\ln y = 0$ for any value of $\nu$ as mentioned above. This property does not hold for $P_{\text{K-PT}}(\ln y)$. The peak of the $\kappa$-deformed Porter-Thomas distribution for less chaotic systems occurs at $\ln y < 0$ and moves towards lower values as the parameter $\kappa$ increases. We show in the next subsection that this is indeed the behavior of physical systems.

### 6.2 Data analysis

Hamoudi, Nazmitdinov and Alhassid [49] calculated the electric quadrupole (E2) and magnetic dipole (M1) transition intensities among the isospin $T = 0, 1$ states of nuclei with mass number 60. They applied the interacting shell model with realistic interaction for $pf$ shell nuclei with a $^{56}$Ni core. It is found that the $B$(E2) transitions are well described by a GOE (Porter-Thomas distribution). However, the statistics for the $B$(M1) transitions is sensitive to $T_z$. The M2 transition operator consists of an isoscalar and isovector components. The $T_z = 1$ nuclei, in which both components contribute, exhibit a Porter-Thomas distribution. In the meanwhile, a significant deviation from the GOE statistics for the $T_z = 0$ nuclei, where the matrix elements are purely isoscalar and relatively weak [50].

We use using the $\kappa$-deformed Porter-Thomas distribution to analyze the reduced M1 transition intensities for both the $T_z = 1$ $^{60}$Co nuclei and $T_z = 0$ $^{60}$Zn nuclei calculated by Hamoudi et al. [49]. These authors sampled a large number of matrix elements for each transition operator,which is equal to $56^2 = 3136$ and $66^2 = 4356$. Figure 8 compares the results of calculations using Eq. (39) with the numerical results of Hamoudi et al. [49] as well as the "best-fit" $\chi^2$ distribution deduced by these authors. The figure clearly shows the advantage of the using the $\kappa$-deformed Porter-Thomas distribution over the $\chi^2$ distribution, at least for this numerical experiment.

# 7 Conclusion

Tsallis' entropy has been considered by several authors as a starting point for constructing a generalization of RMT. However, there are several other generalized entropies. This paper derives $\kappa$-deformed orthogonal and unitary random-matrix ensembles by maximizing the generalized entropy proposed by Kaniadakis under the constraints on normalization of the distribution and the expectation value of $\mathrm{Tr}\big(H^\dagger H\big)$. We obtain a simple formula for the distribution function in the matrix-element space of both of the $\kappa$-deformed GOE and GUE. This leads to new expressions for the NNS distributions of the eigenvalues when $\kappa \neq 0$. The case of $\kappa = 0$ is the one that leads to the Wigner surmise when the entropy is given by the Shannon measure. The high accuracy of Wigner's distribution describing chaotic systems justifies the use of this distribution as a starting point for introducing the $\kappa$ deformations which follow as a consequence of adopting the Kaniadakix entropy. As in the case of Tsallis' entropy, the NNS distribution obtained here for Kaniadakis' entropy has an intermediate shape between the Wigner distribution and the Poissonian that describes generically integrable systems. However, neither of these distributions reaches the Poissonian form. We have explored the possibility to us the NNS distribution obtained in this paper for modelling systems with mixed regular-chaotic dynamics at least when they are not far from the state of chaos. We have tested these expressions by comparing their predictions with the NNS distributions obtained in a number of numerical experiment. We also introduce a $\kappa$ deformation to the Porter-Thomas distribution that describes the transition intensities in chaotic systems. The resulting transition-intensity distribution is found to agree with the results of shell model calculation for a number of nuclei better that the $\chi^2$ distribution, which is often used for this purpose.

# Figure caption

FIG. 1. NNS distributions obtained by using the Kaniadakis statistics. The entropic indices used are $\kappa = 0$ (GOE), $0.3$ and $0.5$. The semi=Poissonian distribution is also plotted to show that it has a peak nearly at the same position as that of the limiting case of the $\kappa$-deformed Wigner surmise (with $\kappa = 1/2$).

FIG. 2. NNS distributions obtained in a point processes on the family of Koch fractals in a 2-dimensional space [35] are shown by histograms. The similarity dimension $d_s$ of the fractals are respectively $d_s = 1.002, 1.060, 1.140, 1.267, 1.375, 1.496, 1.684\,, 1.812$ and 1.988. The continuos curves are calculated by Eq. (29) with values of the entropic index $\kappa$ shown in each window.

FIG. 3. The fractal dimensional $d_s$ corresponding to the NNS distributions shown in Fig. 2 plotted against the best-fit values of entropic index $\kappa$. The continuos line is a parametrization of the empirical values by $d_s(\kappa) = 2.01 - e^{-(\kappa-0.573)^2/(2\times 0.134^2)}$.

FIG. 4. NNS distributions of levels of the elliptical stadium billiards with shape parameters $\gamma = \delta$ [37], fitted to the distributions calculated by Eq. (29) with values of the entropic index $\kappa$ shown in each window.

FIG. 5. NNS distributions for interacting bosons with Hamiltonian $H = H_1 + \lambda H_2$, for various values of $\lambda$ calculated in [41] (a) with 10 bosons in 4 single-particle states and (b) with 10 bosons in 5 single-particle states, fitted to the distributions calculated by Eq. (29) with values of the entropic index $\kappa$ shown in each window. The last histograms in both (a) and (b) are for EGOE(2).

FIG. 6 NNS of $2^+$ of even-even nuclei in different ranges of the $R_{4/2}$ ratio fitted to the distributions calculated by Eq. (29) with values of the entropic index $\kappa$ shown in each window.

FIG. 7. Evolution of the $\kappa$-deformed Porter-Thomas distribution $P_{\text{K-PT}}(\ln y)$ and the $\chi^2$distribution $P_{\chi^2}(\ln y, \nu)$ during the transition from chaotic to regular dynamics. The solid curves, labeled as PT, refer to the Porter-Thomas distribution.

FIG. 8. Nuclear shell-model M1 transition intensities in $A = 60$, calculated by Hamoudi et al. [49], (histograms) compared with the $\kappa$-deformed Porter-Thomas distribution (39) with parameters $\kappa = 0.38,\ 0.61$ and $0.63$, respectively (solid curves) and the $\chi^2$distribution (17) with parameters $\nu = 1,\ 0.64$ and $0.34$, respectively (dashed curves).

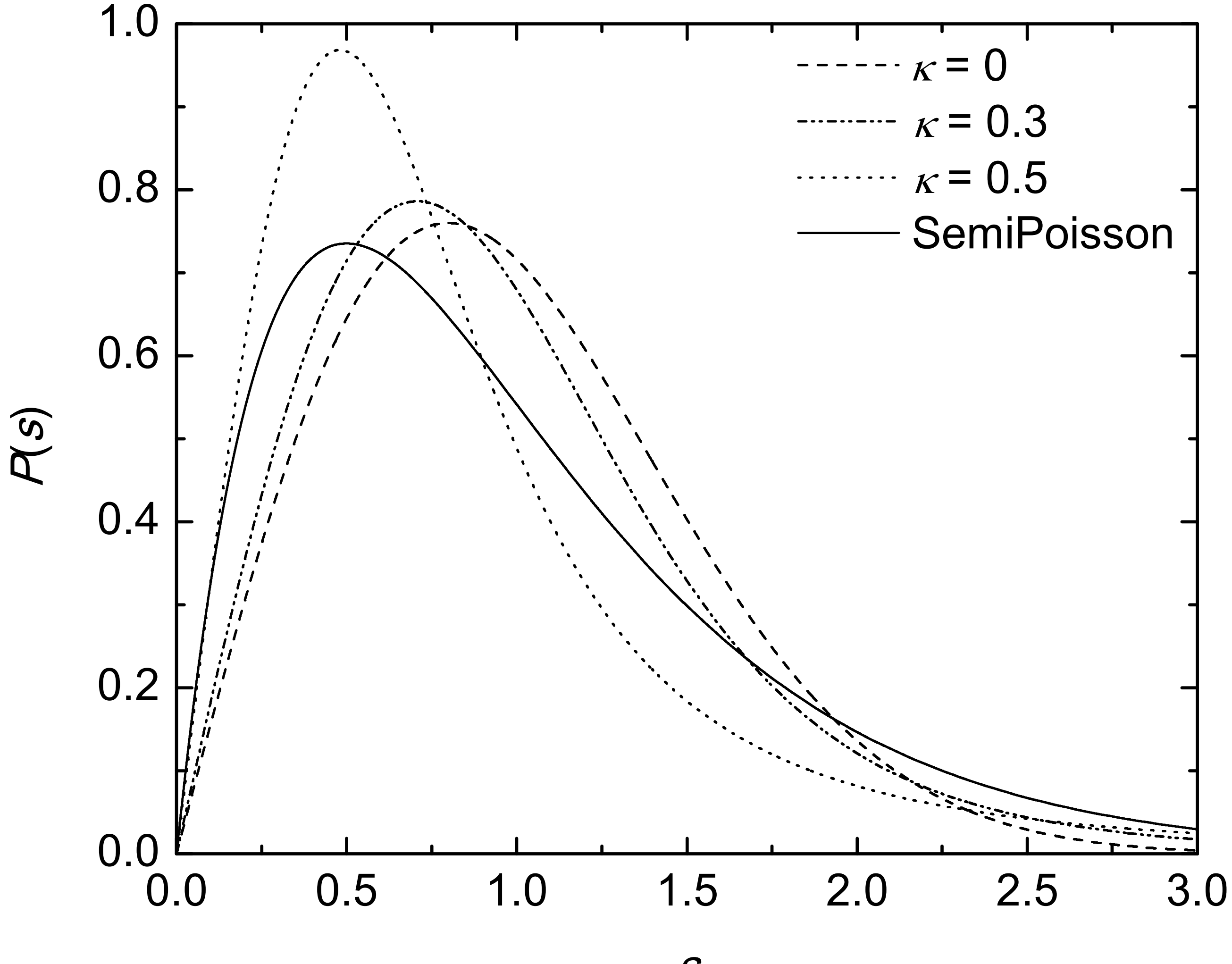

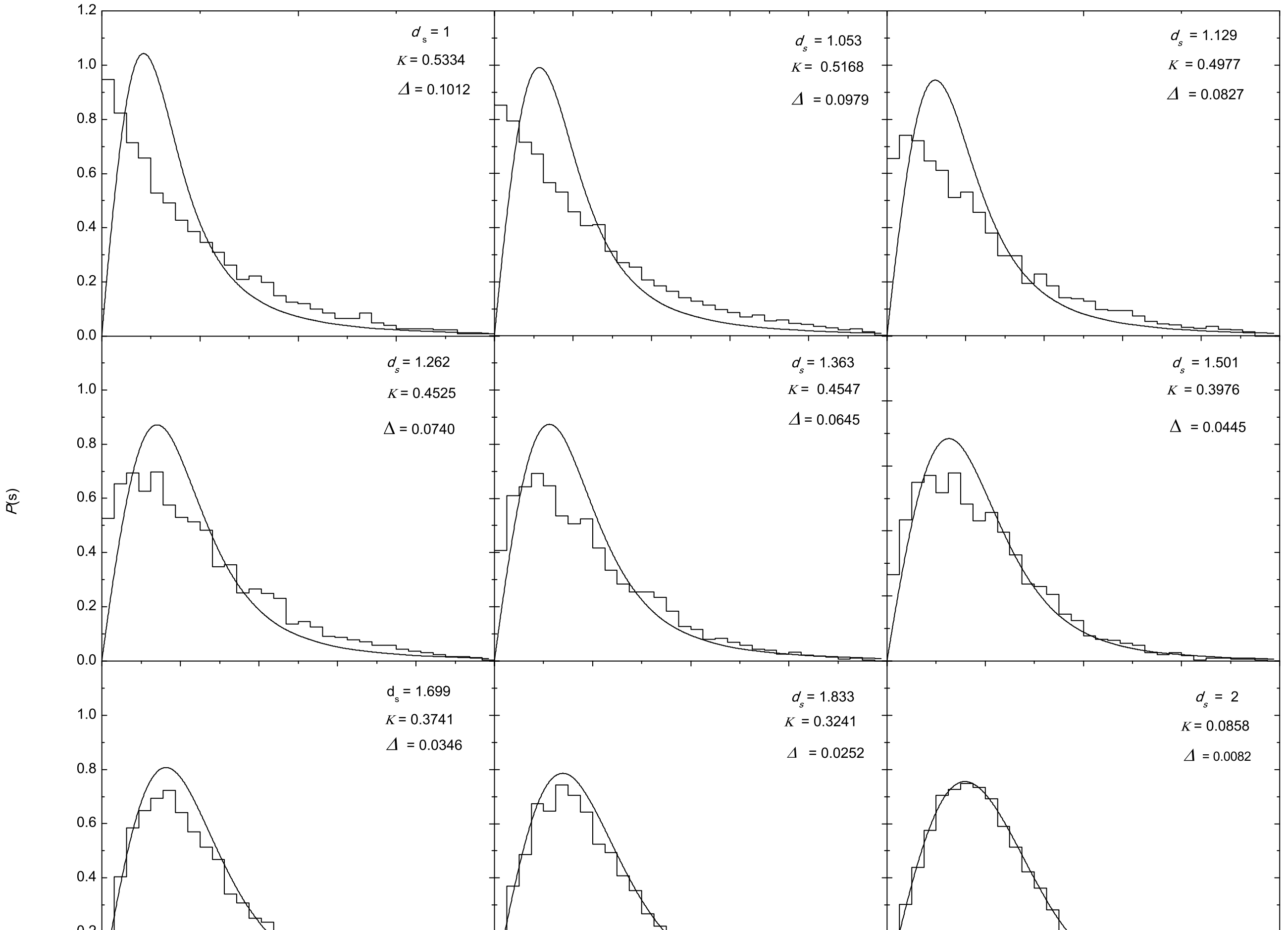

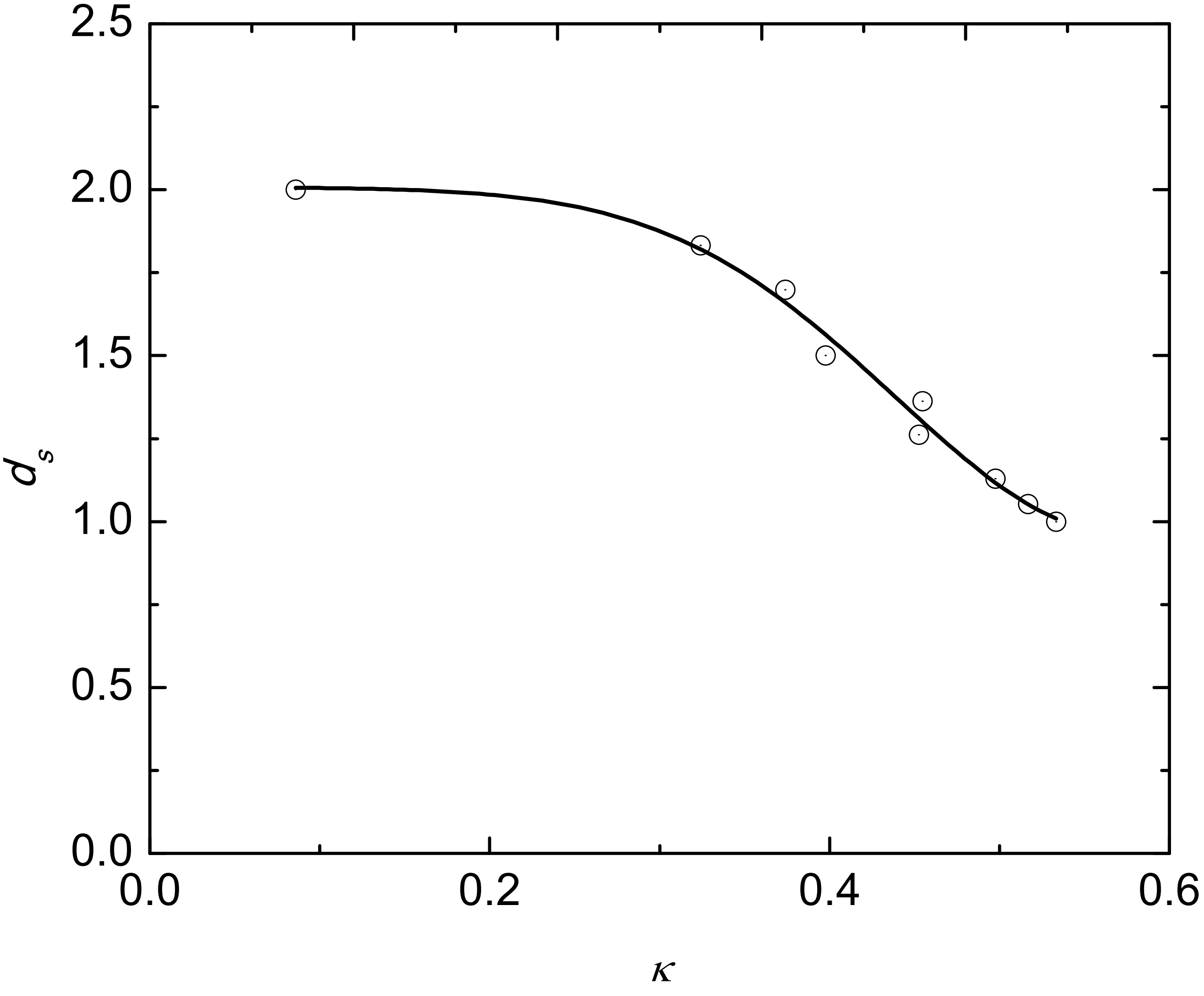

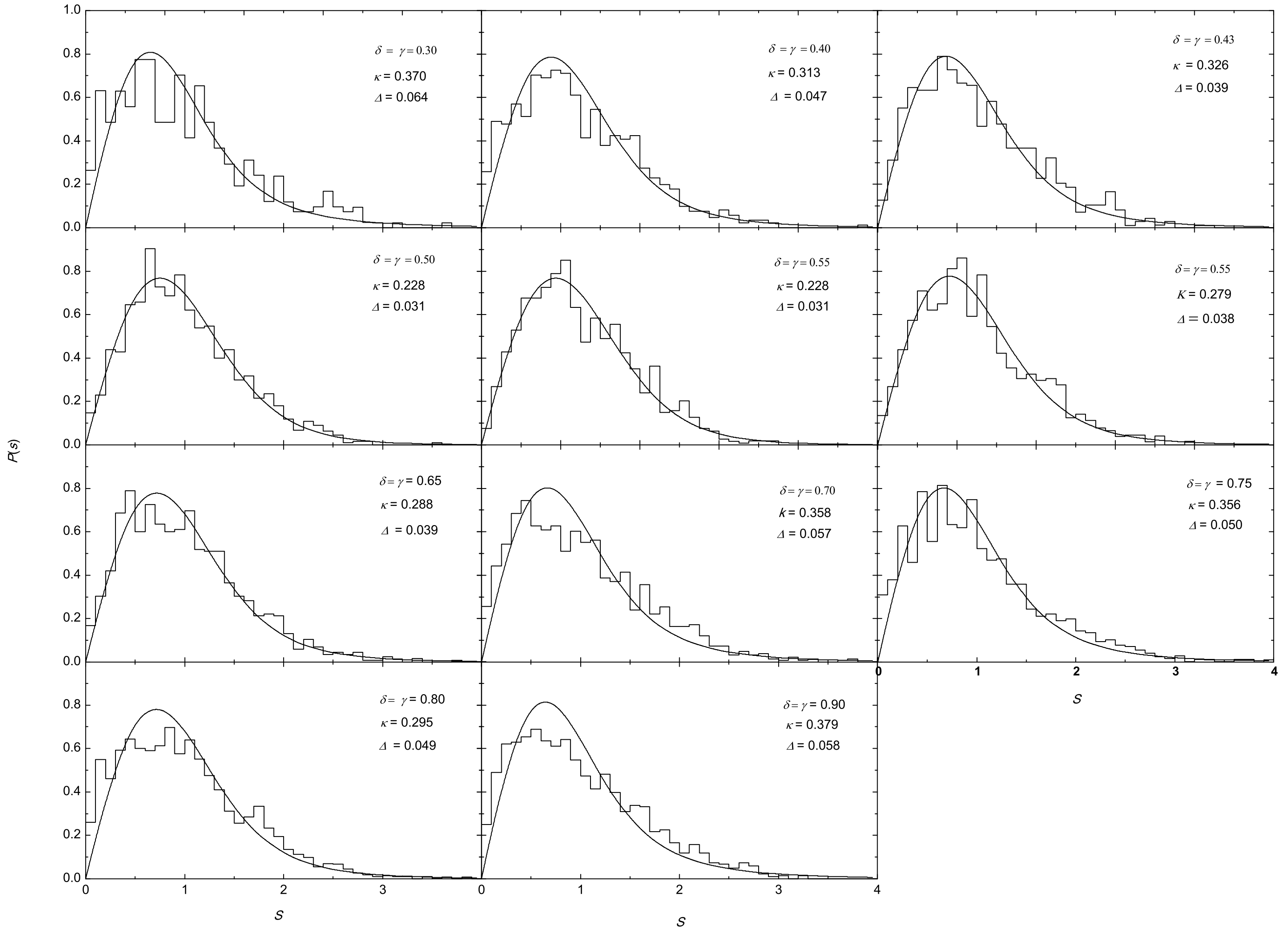

P(s)
S
δ = γ = 0.30
κ = 0.370
Δ = 0.064
δ = γ = 0.40
κ = 0.313
Δ = 0.047
δ = γ = 0.43
κ = 0.326
Δ = 0.039
δ = γ = 0.50
κ = 0.228
Δ = 0.031
δ = γ = 0.55
κ = 0.228
Δ = 0.031
δ = γ = 0.55
K = 0.279
Δ = 0.038
δ = γ = 0.65
κ = 0.288
Δ = 0.039
δ = γ = 0.70
k = 0.358
Δ = 0.057
δ = γ = 0.75
κ = 0.356
Δ = 0.050
δ = γ = 0.80
κ = 0.295
Δ = 0.049
δ = γ = 0.90
κ = 0.379
Δ = 0.058

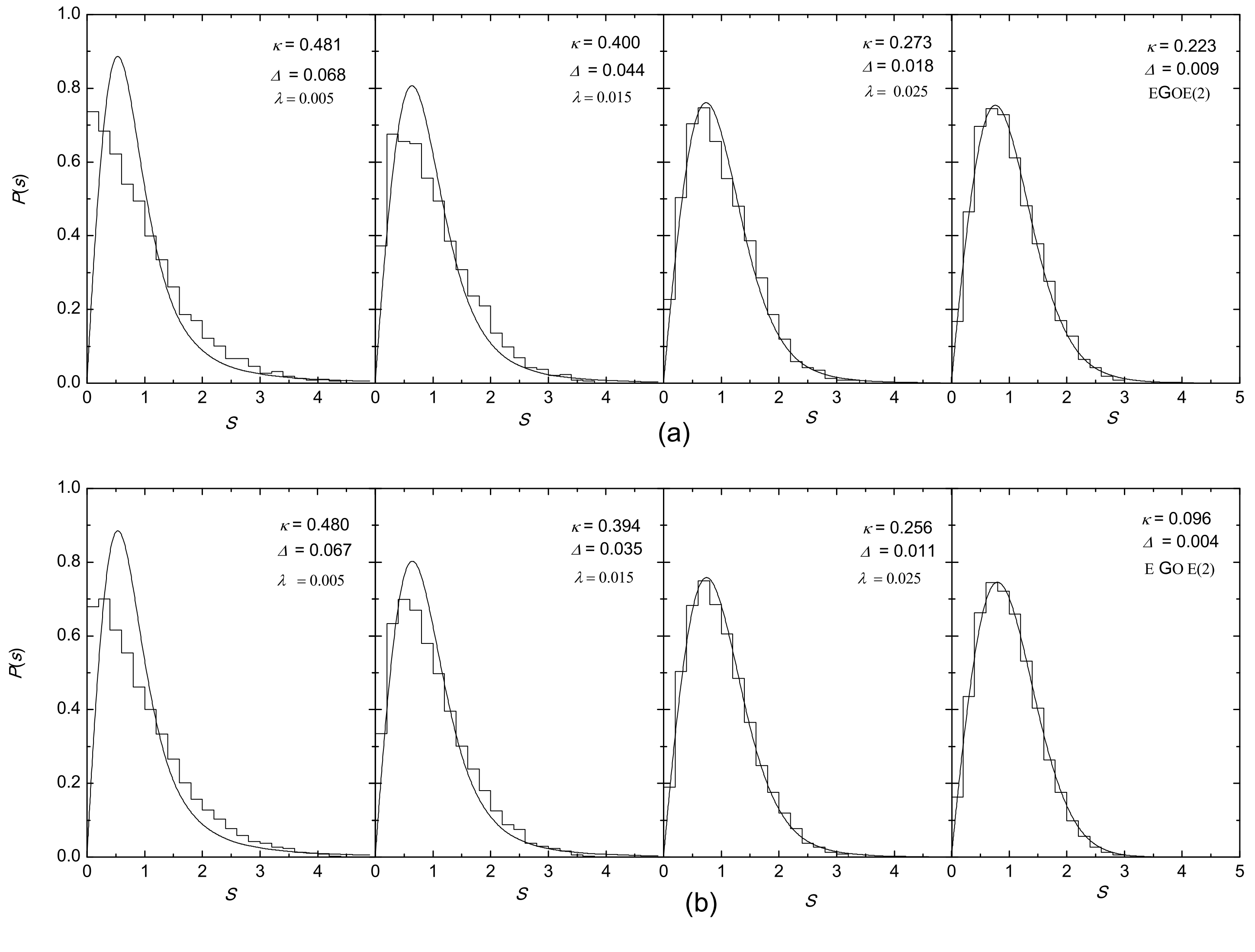

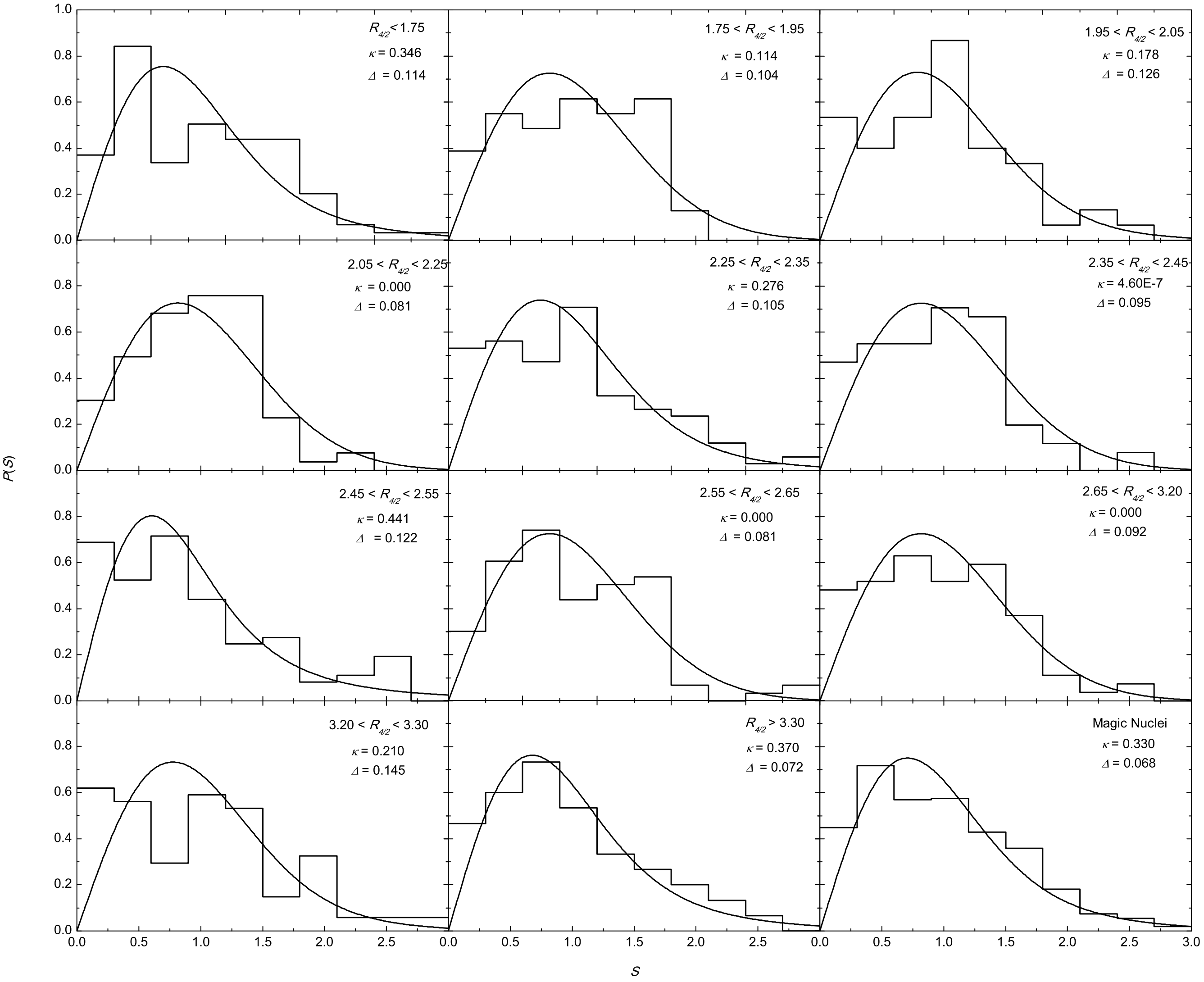
$R_{4/2}$ < 1.75
κ = 0.346
Δ = 0.114
1.75 < $R_{4/2}$ < 1.95
κ = 0.114
Δ = 0.104
1.95 < $R_{4/2}$ < 2.05
κ = 0.178
Δ = 0.126
2.05 < $R_{4/2}$ < 2.25
κ = 0.000
Δ = 0.081
2.25 < $R_{4/2}$ < 2.35
κ = 0.276
Δ = 0.105
2.35 < $R_{4/2}$ < 2.45
κ = 4.60E-7
Δ = 0.095
2.45 < $R_{4/2}$ < 2.55
κ = 0.441
Δ = 0.122
2.55 < $R_{4/2}$ < 2.65
κ = 0.000
Δ = 0.081
2.65 < $R_{4/2}$ < 3.20
κ = 0.000
Δ = 0.092
3.20 < $R_{4/2}$ < 3.30
κ = 0.210
Δ = 0.145
$R_{4/2}$ > 3.30
κ = 0.370
Δ = 0.072
Magic Nuclei
κ = 0.330
Δ = 0.068
P(S)
S

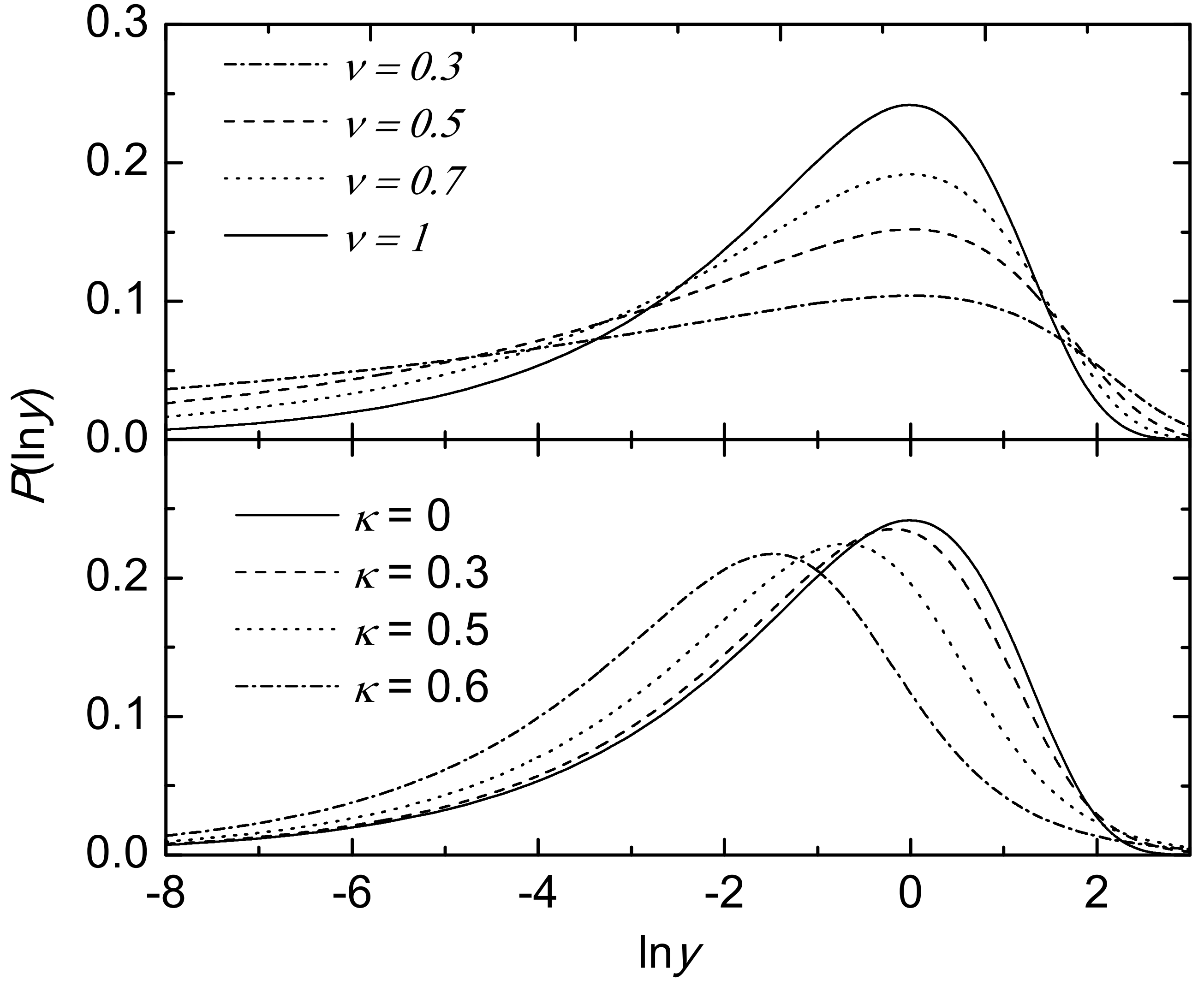

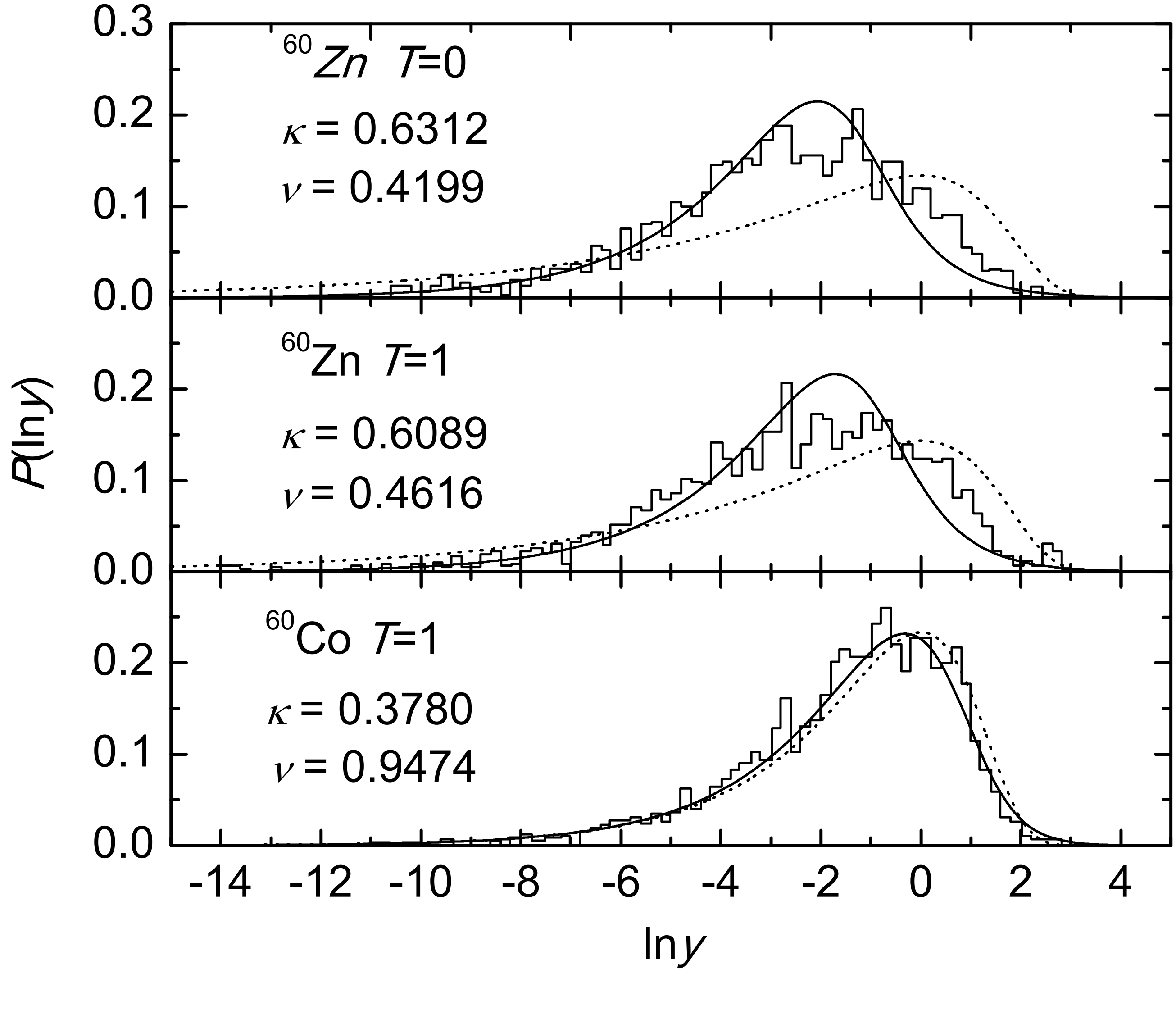